\documentclass[12pt]{iopart}
\usepackage{amsfonts}
\usepackage{amssymb}
\usepackage{amsthm}

\usepackage[dvips]{epsfig}

\begin{document}
\title{Dobi\'nski-type relations and the Log-normal distribution}
\author{P Blasiak$^{\dag\ddag}$,
 K A Penson$^{\dag}$ and A I Solomon$^{\dag}$}
\address{\ \linebreak$^{\dag}$Universit\'{e} Pierre et Marie Curie,
Laboratoire   de  Physique   Th\'{e}orique  des  Liquides, CNRS UMR 7600\linebreak
Tour 16, $5^{i\grave{e}me}$ \'{e}tage, 4, place Jussieu, F 75252 Paris
Cedex 05, France\linebreak}
\address{$^{\ddag}$H. Niewodnicza{\'n}ski Institute of Nuclear Physics,\linebreak
ul.Eliasza Radzikowskiego 152, PL 31-342 Krak{\'o}w, Poland\linebreak}
\eads{\linebreak\mailto{blasiak@lptl.jussieu.fr},
\mailto{penson@lptl.jussieu.fr}, \mailto{a.i.solomon@open.ac.uk}\linebreak}

\begin{abstract}
\linebreak
We consider sequences of generalized Bell numbers $B(n),\ n=0,1,\ldots$
which can be represented by Dobi\'nski-type summation formulas,
i.e. $B(n)=\frac{1}{C}\sum_{k=0}^\infty \frac{[P(k)]^n}{D(k)}$, with
$P(k)$ a polynomial, $D(k)$ a function of $k$ and $C=const$. They
include the standard Bell numbers ($P(k)=k$, $D(k)=k!$, $C=e$), their
generalizations $B_{r,r}(n),\ r=2,3,\ldots$ appearing in the normal ordering of
powers of boson monomials ($P(k)=\frac{(k+r)!}{k!}$, $D(k)=k!$, $C=e$), variants of ``ordered'' Bell numbers $B_o^{(p)}(n)$ ($P(k)=k$,
$D(k)=(\frac{p+1}{p})^k$, $C=1+p$, p=1,2\ldots),
 etc. We demonstrate that for $\alpha,
\beta, \gamma, t$ positive integers ($\alpha, t\neq 0$), $\left[B(\alpha n^2+\beta n+\gamma)\right]^t$
is the $n$-th moment of a positive function on $(0,\infty)$ which is a
weighted infinite sum of log-normal distributions.

\end{abstract}

\maketitle

In a recent investigation \cite{BPS} we analysed  sequences of
integers which appear in the process of normal ordering of powers
of monomials of boson creation $a^\dag$ and annihilation $a$
operators, satisfying  the commutation rule $[a,a^\dag]=1$. For
$r,s$ integers such that $r\geq s$, we define the generalized
Stirling numbers of the second kind $S_{r,s}(n,k)$ as
\begin{eqnarray}
    \left[(a^\dag)^ra^s\right]^n=(a^\dag)^{n(r-s)}\sum_{k=s}^{ns}S_{r,s}(n,k)(a^\dag)^ka^k
\end{eqnarray}
and the corresponding Bell numbers $B_{r,s}(n)$ as
\begin{eqnarray}
    B_{r,s}(n)=\sum_{k=s}^{ns}S_{r,s}(n,k).
\end{eqnarray}
In \cite{BPS} explicit and exact expressions for $S_{r,s}(n,k)$
and $B_{r,s}(n)$ were found. In a parallel study \cite{PS} it was
demonstrated that $B_{r,s}(n)$ can be considered as the $n$-th
moment of a probability distribution on the positive half-axis. In
addition, for every pair $(r,s)$ the corresponding distribution
can be explicitly written down. These distributions  constitute
the solutions of a family of Stieltjes moment problems, with
$B_{r,s}(n)$ as moments. Of particular interest to us are the
sequences with $r=s$, for which the following representation as an
infinite series has been obtained:
\begin{eqnarray}
B_{r,r}(n)&=&\frac{1}{e} \sum_{k=0}^\infty \frac{1}{k!}\left[\frac{(k+r)!}{k!}\right]^{n-1}\label{A}\\
&=&\frac{1}{e} \sum_{k=0}^\infty\frac{\left[k(k+1)\ldots(k+r-1)\right]^n}{(k+r-1)!},\ \ \ \ \
n>0.\label{B}
\end{eqnarray}
Eqs.(\ref{A}) and (\ref{B}) are generalizations of
the celebrated Dobi\'nski formula ($r=1$) \cite{Comtet}:
\begin{eqnarray}
B_{1,1}(n)=\frac{1}{e}\sum_{k=0}^\infty \frac{k^n}{k!},\ \ \ \ n\geq0,\label{C}
\end{eqnarray}
which expresses the conventional Bell numbers $B_{1,1}(n)$ as a
rapidly convergent series. Its simplicity has inspired
combinatorialists such as G.-C. Rota \cite{Rota} and H.S. Wilf
\cite{Wilf}. Eq.(\ref{C}) has far-reaching implications in
the theory of stochastic processes \cite{ConstSav}, \cite{Pitman},
\cite{Const}.

The probability distribution whose $n$-th moment is $B_{r,r}(n)$
is an infinite ensemble of weighted Dirac delta functions located
at a specific set of integers (a so-called {\em Dirac comb}):
\begin{eqnarray}
B_{r,r}(n)=\int_0^\infty x^n\left\{\frac{1}{e}\sum_{k=0}^\infty\frac{\delta
(x-k(k+1)\ldots(k+r-1))}{(k+r-1)!}\right\}dx,\ \  n\geq0.\label{D}
\end{eqnarray}

For $r=1$ the discrete distribution of Eq.(\ref{D}) is the weight
function for the orthogonality relation for Charlier polynomials \cite{Koek}.
In contrast we emphasize that for $r\neq s$ the $B_{r,s}(n)$ are moments
of continuous distributions \cite{PS}.
\begin{figure}

\resizebox{14cm}{!}{\includegraphics{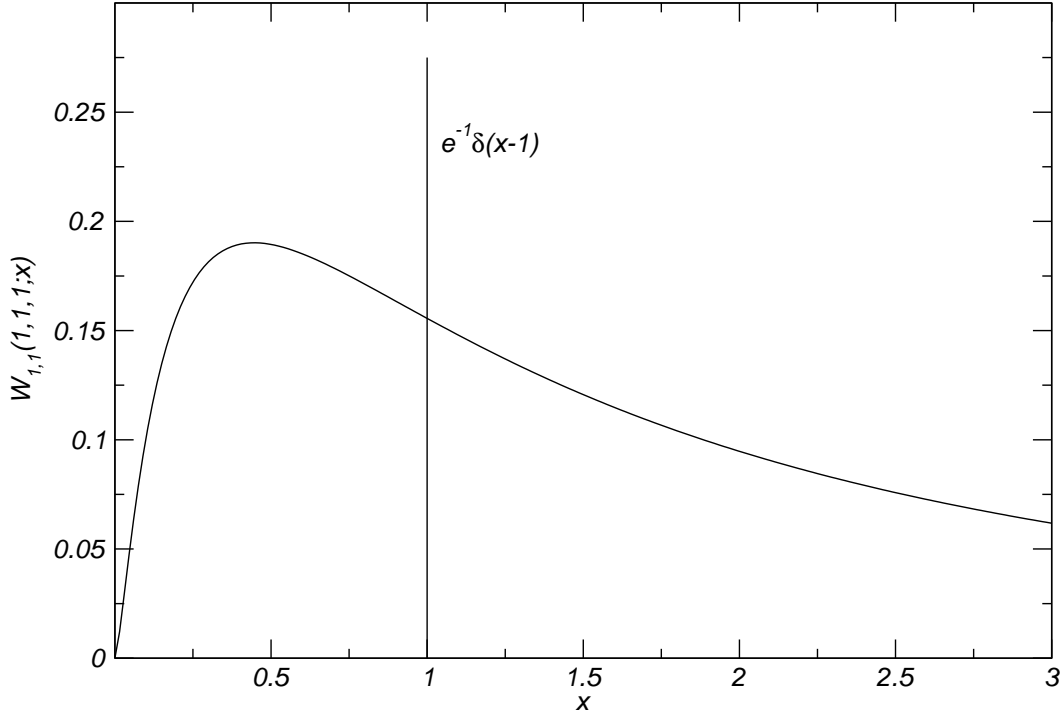}}
\caption{\label{1}Weight function $W_{1,1}(1,1,1;x)$, see Eq.(\ref{H}).}
\end{figure}

In this note we wish to point out an intimate relation between the
formulas of Eqs. (\ref{A}), (\ref{B}), (\ref{C}) and the
log-normal distribution \cite{Crow}, \cite{Yor}:
\begin{eqnarray}
P_{\sigma,\mu}(x)=\frac{1}{\sqrt{2\pi}\sigma x}e^{-\frac{(\ln
(x)-\mu)^2}{2\sigma^2}},\ \ \ \ \ \ \ \   x\geq0,\ \ \ \
\sigma,\mu>0. \label{E}
\end{eqnarray}
First we quote the standard expression for its $n$-th moment:
\begin{eqnarray}
M_n=\int_0^\infty x^nP_{\sigma,\mu}(x)dx=e^{n\left(\mu+n\frac{\sigma^2}{2}\right)},\ \ n\geq 0,
\end{eqnarray}
which can be reparametrized for $k>1$ as
\begin{eqnarray}
M_n=k^{\alpha n^2+\beta n},\label{M}
\end{eqnarray}
with
\begin{eqnarray}
\mu &=&\beta \ln (k),\\
\sigma &=&\sqrt{2\alpha \ln (k)}>0.
\end{eqnarray}
Given three integers $\alpha,\beta,\gamma$ (where $\alpha>0$), we
wish  to find a weight function $W_{1,1}(\alpha,\beta,\gamma;x)>0$
such that
\begin{eqnarray}
B_{1,1}(\alpha n^2+\beta n+\gamma)=\int_0^\infty x^nW_{1,1}(\alpha,\beta,\gamma;x)dx.\label{F}
\end{eqnarray}
Eqs.(\ref{C}), (\ref{E}) and (\ref{M}) provide an immediate
solution:
\begin{eqnarray}
W_{1,1}(\alpha,\beta,\gamma;x)=\frac{1}{e}\left[\delta(x-1)+\sum_{k=2}^\infty \frac{k^\gamma
\exp\left(-\frac{(\ln (x)-\beta \ln (k))^2}{4\alpha \ln (k)}\right)}{2xk!\sqrt{\pi\alpha \ln (k)}}\right],\label{H}
\end{eqnarray}
which is an \emph {infinite} sum of weighted log-normal
distributions supplemented by a single Dirac peak of weight
$e^{-1}$ located at $x=1$. Thus it is a \emph {superposition} of
discrete and continuous distributions. Virtually the same approach
can be adopted for the sequences $B_{r,r}(n)$, $r>1$. In this case
the $k=1$ term in the numerator of Eq.(\ref{A}) is larger than one
and so  there will be no Dirac peak in the formula.  Then the
function $W_{r,r}(\alpha,\beta,\gamma;x)>0$ defined by
($\alpha,\beta,\gamma$ integers, $\alpha,\gamma >0$):
\begin{eqnarray}
B_{r,r}(\alpha n^2+\beta n+\gamma)=\int_0^\infty x^nW_{r,r}(\alpha,\beta,\gamma;x)dx,\label{G}
\end{eqnarray}
is a purely \emph {continuous} probability distribution given again by an
infinite sum of weighted log-normal distributions:
\begin{eqnarray}
W_{r,r}(\alpha,\beta,\gamma;x)=\frac{1}{e}\sum_{k=0}^\infty
\frac{\left[\frac{(k+r)!}{k!}\right]^{\gamma -1}
\exp\left(-\frac{\left[\ln (x)-\beta \ln
\left(\frac{(k+r)!}{k!}\right)\right]^2}{4\alpha \ln
\left[\frac{(k+r)!}{k!}\right]}\right)}{2xk!\sqrt{\pi\alpha \ln
\left[\frac{(k+r)!}{k!}\right]}}.\label{I}
\end{eqnarray}
\begin{figure}

\resizebox{14cm}{!}{\includegraphics{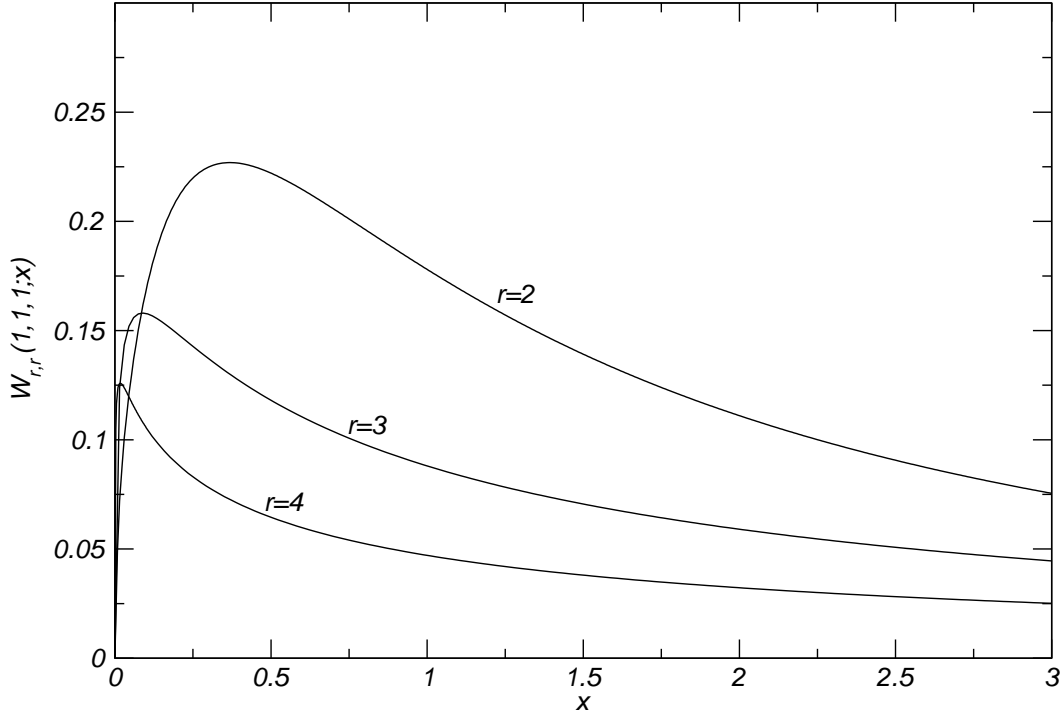}}
\caption{\label{2}Weight functions $W_{r,r}(1,1,1;x)$ for
$r=2,3,4$.}
\end{figure}
The solutions of the moment problems of Eqs.(\ref{M}), (\ref{F}) and
(\ref{G}) are not unique. More general solutions may be obtained by
the method of the inverse Mellin transform, see \cite{Sixdeniers}.

Several other types of combinatorial sequences have properties
exemplified by Eqs.(\ref{F}) and (\ref{G}).
We quote for example the so-called ``ordered'' Bell numbers $B_o(n)$
defined as \cite{Wilf}:
\begin{eqnarray}
B_o(n)=\sum_{k=1}^n S(n,k)k!,
\end{eqnarray}
where $S(n,k)$ are the Stirling numbers of the second kind,
$S_{1,1}(n,k)$ in our notation. These ordered Bell numbers satisfy
the following Dobi\'nski-type relation:
\begin{eqnarray}
B_o(n) = \frac{1}{2}\sum_{k=0}^\infty \frac{k^n}{2^k}, \label{K}
\end{eqnarray}
from which a formula analogous to Eq.(\ref{H}) readily
follows.
A more general identity of type (\ref{K}) is \cite{Weisstein}:
\begin{eqnarray}
B_o^{(p)}(n)=\frac{1}{p+1}\sum_{k=1}^\infty
k^n\left(\frac{p}{p+1}\right)^k=\sum_{k=0}^nS(n,k)k!\ p^k,\ \ \ \ \
p=2,3,\ldots.\label{P}
\end{eqnarray}

We will not discuss other types of sequences but rather observe
that the relations of Eqs. (\ref{A}), (\ref{B}), (\ref{C}),
(\ref{K}),(\ref{P}) naturally imply that any power of these
numbers also satisfies  a Dobi\'nski-type relation. As an example
we give  explicitly the simplest case of Eq.(\ref{C}). For integer
$t>0$:
\begin{eqnarray}
\left[B_{1,1}(n)\right]^t=\frac{1}{e^t}\sum_{k_1,k_2,\ldots k_t=0}^\infty \frac{(k_1k_2\ldots k_t)^n}{k_1!k_2!\ldots k_t!},\label{J}
\end{eqnarray}
with correspondingly more complicated formulas of a similar nature
for powers of $B_{r,r}(n)$, $B_o(n)$ and $B_o^{(p)}(n)$. For
combinatorial applications of Eq.(\ref{J}) see \cite{Pittel},
\cite{Canfield} and \cite{Bender}.

We conclude that for any sequences of the type
 $B(n)$ specified above $\left[B(\alpha n^2+\beta n+\gamma)\right]^t$ is
always given as an $n$-th moment of a positive function on
$(0,\infty)$ expressible by sums of weighted log-normal
distributions. We illustrate such a function for $B_{1,1}(n)$ in
Fig.(\ref{1}). The application to $B_{r,r}(n)$ for $r=2,3,4$ is
presented in Fig.(\ref{2}).  The area under every curve is equal
to 1 on extrapolating to large $x$ (not displayed).
 In both examples we have
chosen $\alpha=\beta=\gamma=1$. Observe the exceedingly slow
decrease of these probabilities for $x\to \infty$. This is
confirmed by the fact that the moment sequences $\left[B(\alpha
n^2+\beta n+\gamma)\right]^t$ are extremely rapidly increasing. In
the simplest case $\alpha=\beta=\gamma=t=1$ we find
$B_{1,1}(n^2+n+1)=1,5,877,27644437,474869816156751$ for
$n=0\ldots4$.

The circumstance that we can determine the positive solutions of
the Stieltjes moment problem with both $B(n)$ (discrete
distribution) and $\left[B(\alpha n^2+\beta n+\gamma)\right]^t$
(continuous distribution) is a very specific consequence  of the
existence of  Dobi\'nski-type expansions. To our knowledge it has
no equivalent in standard solutions of the moment problem. For
instance, if the moments are $n!$ the solution $e^{-x}$ does not
give any indication as how one might obtain the solution for the
moments equal to $(n^2)!$ .

The strict positivity of $W_{r,r}(\alpha,\beta,\gamma;x)$, for
$r=1,2\ldots$, suggests their use in the construction of coherent
states, which satisfy the {\em resolution of identity} property
\cite{SPS}, \cite{KPS}, \cite{Quesne1}, \cite{Quesne2}.
 This can be done by the substitution
$n!\to B_{r,r}(\alpha n^2+\beta n+\gamma)$ in the definition of standard coherent states. More
precisely for a complete and orthonormal set of wave functions $|n\rangle$ such that $\langle
n|n'\rangle =\delta_{n,n'}$ and complex $z$ we define the normalized coherent
state as
\begin{eqnarray}
|z;\alpha,\beta,\gamma\rangle = \frac{1}{{\mathcal N}^{1/2}(\alpha,\beta,\gamma;|z|^2)}\sum_{n=0}^\infty
\frac{z^n}{\sqrt{B_{r,r}(\alpha n^2+\beta n+\gamma)}}|n\rangle,\label{Z}
\end{eqnarray}
with the normalization
\begin{eqnarray}
{\mathcal N}(\alpha,\beta,\gamma;x)=\sum_{n=0}^\infty
\frac{x^n}{B_{r,r}(\alpha n^2+\beta n+\gamma)},
\end{eqnarray}
which is a rapidly converging function of $x$ for $0\leq x<
\infty$, $x=|z|^2$. Then, using the procedure of \cite{KPS} we
can demonstrate that the states of Eq.(\ref{Z}) along with
Eq.(\ref{G}) automatically
satisfy the resolution of unity
\begin{eqnarray}
{\int\int}_{\mathbb{C}}d^2z|z;\alpha,\beta,\gamma\rangle \tilde{W}_{r,r}(\alpha,\beta,\gamma;|z|^2)\langle
z;\alpha,\beta,\gamma|=I=\sum_{n=0}^\infty |n\rangle\langle n|,
\end{eqnarray}
with
\begin{eqnarray}
 W_{r,r}(\alpha,\beta,\gamma;|z|^2)=\pi\frac{\tilde{W}_{r,r}(\alpha,\beta,\gamma;|z|^2)}{{\mathcal N}(\alpha,\beta,\gamma;|z|^2)}.
\end{eqnarray}
We are currently investigating the quantum-optical properties of
states defined in Eq.(\ref{Z}).

We close by quoting from \cite{Pitman} that, ``the idea of
representing the combinatorially defined numbers by an infinite
sum or an integral, typically with a probabilistic interpretation,
has proved to be a very fruitful one''. In our particular case it
has allowed us to reveal quite an unexpected relation between the
Dobi\'nski-type summation relations, which by themselves are
reflections of boson statistics, and the log-normal distribution.

\ack

We thank D. Barsky, G. Duchamp, L. Haddad, A. Horzela and M.Yor for
numerous fruitful discussions. N.J.A. Sloane's Encyclopedia of
Integer Sequences (http://www.research.att.com/{\textasciitilde}njas/sequences) was an
essential help in analyzing the properties of sequences we have
encountered in this work.

\end{document}